\documentclass[aps,prd,preprint,groupedaddress]{revtex4-2}

\usepackage{slashed}
\usepackage{graphicx,color}
\usepackage{amsmath}
\usepackage{amssymb}
\begin{document}

\preprint{{SI-HEP-2022-33, P3H-22-110}}

\title{On the contribution of the electromagnetic dipole operator ${\cal O}_7$ \\ to the $\bar B_s \to \mu^+\mu^-$ decay amplitude}

\author{Thorsten Feldmann}
\email[]{thorsten.feldmann@uni-siegen.de}

\author{Nico Gubernari}
\email[]{nicogubernari@gmail.com}

\author{Tobias Huber}
\email[]{huber@physik.uni-siegen.de} 

\author{Nicolas Seitz}
\email[]{nicolas.seitz@uni-siegen.de}

\affiliation{Theoretische Physik 1, Center for Particle Physics Siegen (CPPS), Universit\"at Siegen, Walter-Flex-Stra{\ss}e 3, D-57068 Siegen, Germany}

\date{\today}

\begin{abstract}

We construct a factorization theorem that allows
to systematically include QCD corrections to the contribution of the electromagnetic dipole operator in the effective weak Hamiltonian to the $\bar B_s \to \mu^+\mu^-$ decay amplitude.
We first rederive the known result for the leading-order QED box diagram, which features a double-logarithmic enhancement associated to the different rapidities of the light quark in the $\bar B_s$ meson and the energetic muons in the final state. We provide a detailed analysis of the cancellation of the related endpoint divergences appearing in individual momentum regions, and show how the rapidity logarithms can be isolated by suitable subtractions applied to the corresponding bare factorization theorem.
This allows us to include in a straightforward manner the QCD corrections arising from the renormalization-group running of the hard matching coefficient of the electromagnetic dipole operator in soft-collinear effective theory, the hard-collinear scattering kernel, and the $B_s$-meson distribution amplitude. Focusing on the contribution from the double endpoint logarithms, we derive a compact formula that resums the leading-logarithmic QCD corrections.
\end{abstract}

\maketitle

\clearpage 

\tableofcontents

\clearpage 

\section{Introduction}

The rare decay $\bar B_s \to \mu^+\mu^-$ mediated by flavour-changing neutral current $b\to s$ transitions represents one of the golden channels to test the flavour sector of the Standard Model (SM).
Precision measurements of its decay rate are
performed at current flavour experiments \cite{LHCb:2012skj,LHCb:2013vgu,CMS:2013dcn,CMS:2014xfa,ATLAS:2016rxw,LHCb:2017rmj,ATLAS:2018cur,CMS:2019bbr,CMS-PAS-BPH-20-003,LHCb:2021vsc,CMS-PAS-BPH-21-006,LHCb:2021awg}.
Precise theoretical predictions include higher-order QCD and electroweak corrections (see e.g.~Refs.~\cite{DeBruyn:2012wk,Buras:2012ru,Buras:2013uqa,Bobeth:2013uxa}, for reviews see e.g.~Refs.~\cite{LHCb:2012myk,Belle-II:2018jsg} and references therein).
Comparing experimental precision measurements and theoretical predictions thus allow us to search for indirect effects of physics beyond the SM.
As it has been pointed out in Refs.~\cite{Beneke:2017vpq,Beneke:2019slt}, at the level of the intended precision, theoretical calculations also have to take into account corrections from non-local QED effects. In particular, the exchange of an additional photon between the light degrees of freedom in the $\bar B_s$-meson and the final-state muons leads to a power-enhancement compared to the leading contribution from the local semi-leptonic operator ${\cal O}_{10}$. 
Besides the phenomenological importance, these effects are of particular theoretical relevance for two reasons. First, QED corrections to exclusive decays of heavy quarks at large recoil are conceptually interesting, and the corresponding generalization of the QCD factorization approach is non-trivial \cite{Beneke:2020vnb,Beneke:2021jhp,Beneke:2021pkl,Beneke:2022msp}.
Second, the contribution of the electromagnetic dipole operator ${\cal O}_7$ features a double-logarithmic enhancement that results from endpoint configurations of the light-cone momentum fractions of the intermediate muon propagator with respect to the external muons. Analysing the relevant QED box diagram with the method of momentum regions then leads to endpoint-divergent convolution integrals, which makes the formulation of factorization theorems and the renormalisation-group improvement for this contribution difficult and non-standard
\footnote{On the other hand, it has been shown in Ref.~\cite{Beneke:2019slt} that QCD/QED factorization theorems for the contributions of the semi-leptonic operators ${\cal O}_9$ and ${\cal O}_{10}$ can be
obtained in the framework of soft-collinear effective theory without complications from endpoint divergences.}.
The systematic understanding of endpoint logarithms in the context of factorization theorems and effective-field-theory methods is currently a very active field of research. While the analysis of endpoint dynamics and rapidity logarithms for non-perturbative setups, like exclusive charmless $b$-quark decays 
\cite{Beneke:2003zv}, is notoriously difficult \footnote{Recent attempts to understand QCD factorization in $B_c \to \eta_c$ transitions in Ref.~\cite{Boer:2018mgl} indicate that a consistent treatment of endpoint divergences in exclusive charmless $B$-meson decays is more subtle than anticipated in Ref.~\cite{Lu:2022fgz}.}, progress has 
recently been made for a number of perturbative examples, including bottom-induced $h\to \gamma\gamma$ decay \cite{Liu:2019oav,Liu:2020wbn}, off-diagonal gluon thrust \cite{Beneke:2022obx}, or
muon-electron backward scattering \cite{Bell:2022ott}.
In this context, 
the ${\cal O}_7$ contribution to the $\bar B_s \to \mu^+\mu^-$ decay amplitude offers another example to study
rapidity logarithms in hard exclusive transitions. In this case the origin of the endpoint double logarithms resides on the muon side and can thus be understood in QED perturbation theory, while the dynamics of the strange-quark is embedded in the exclusive hadronic $B \to$ vacuum transition.

The aim of this work is to provide a QCD factorization theorem for the ${\cal O}_7$
contribution to the $\bar B_s \to \mu^+\mu^-$ decay amplitude. To this end, in the following section~\ref{sec:LO}, we
first rederive the result from Ref.~\cite{Beneke:2017vpq} by adding up the relevant momentum regions \cite{Beneke:1997zp,Smirnov:1998vk} 
of the QED box diagram. Depending on the choice for the additional regulator that is needed to render the convolution integrals finite, we find different relevant momentum regions. In particular, we show that the double-logarithmic term in the final result can be isolated from a certain momentum configuration, where the photon propagators become eikonal and the intermediate muon propagator goes on-shell
(see e.g.\ Ref.~\cite{Liu:2018czl} and references therein). 
The results from the analysis of the leading QED box diagram can be generalised to obtain a bare factorisation theorem that takes into account additional QCD corrections, depending on the applied regulators. By means of suitable subtractions on the basis of refactorisation conditions, we construct a factorisation theorem with endpoint-convergent convolutions where the additional analytic regulator can be dropped and the $1/\epsilon$ divergences from 
dimensional regularization are manifest.

 On the basis of this result, it is a rather straightforward task to implement the leading-logarithmic QCD corrections to the $b \to s\gamma$ vertex, the jet function describing the exchange of a hard-collinear strange quark and the light-cone distribution amplitude (LCDA) of the $B_s$-meson. This is worked out in detail in Section~\ref{sec:LL}. On the basis of an explicit and systematic parameterization for the LCDA we also provide a relatively compact formula that resums the leading-logarithmic QCD effects in renormalization-group (RG) improved perturbation theory. A brief numerical analysis shows that the effect of the leading-logarithmic QCD corrections can be of the order of $10-30$\% relative to the ${\cal O}_7$ contribution at fixed-order ${\cal O}(\alpha_s^0)$, depending on the shape of the $B_s$-meson LCDA.
It should, however, be noted that the overall effect of the electromagnetic dipole operator is small because of the associated small Wilson coefficient
$C_7$, such that our result only has
a marginal phenomenological impact on the $\bar B_s \to \mu^+\mu^-$ decay rate.
We conclude this paper with a short summary. A detailed description of the subtractions that are required to get rid of endpoint-divergent convolution integrals in the bare factorisation theorem is provided in the appendix.

\section{The leading-order QED box diagram}
\label{sec:LO}

\begin{figure}[t] 
\begin{center}
	\includegraphics[width=0.60\textwidth]{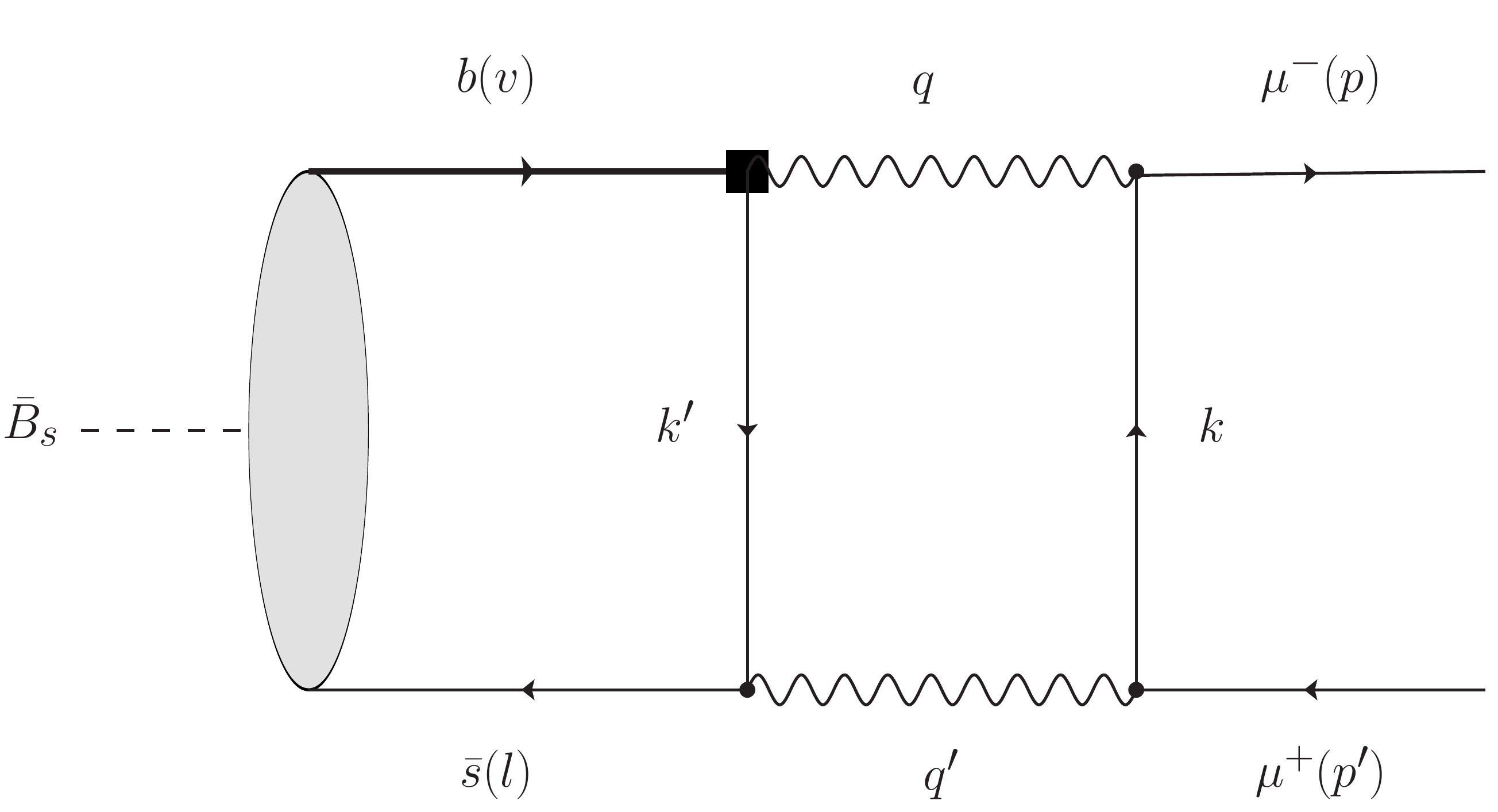}
\end{center}
\caption{\label{fig:box} QED box diagram describing the leading contribution of the operator ${\cal O}_7$ (indicated by the black square) to the $\bar B_s \to \mu^+\mu^-$ decay amplitude. The analogous diagram with the r{\^o}le of $\mu^+$ and $\mu^-$ interchanged gives the same result and is not shown.}
\end{figure}

Our starting point is the non-local matrix element
\begin{equation}
     \langle 0 | \int\! d^4x \, T \{j^\mu_{\rm{em}}(x) , {\cal H}_{\rm{eff}}(0)\}|\bar B_s \rangle \label{eq:nonlocal}
\end{equation}
of the time-ordered product of the quark electromagnetic current $j^\mu_{\rm{em}}(x) = \sum_q Q_q \, \bar q(x) \gamma^\mu q(x)$ and the effective weak Hamiltonian ${\cal H}_{\rm{eff}}$, of which we adopt the convention of Ref.~\cite{Chetyrkin:1996vx} and of which the operator $${\cal O}_7= \frac{e}{16\pi^2} \, m_b \, (\bar s_L \sigma^{\mu\nu}b_R) F_{\mu\nu}$$ is in our focus in this article.
The leading-order contribution of ${\cal O}_7$ to the $\bar B_s \to \mu^+\mu^-$ decay amplitude via the non-local matrix element~(\ref{eq:nonlocal}) is diagrammatically shown in Fig.~\ref{fig:box}, where an additional virtual photon connects the spectator quark of the $\bar B_s$ meson to one of the final-state leptons. The contribution of this diagram (and the corresponding crossed one with the r{\^o}le of the two muons interchanged, which turns out to give the identical contribution) to the $\bar B_s\to\mu^+\mu^-$ decay amplitude has first been calculated in Ref.~\cite{Beneke:2017vpq} and can be written as 
\begin{eqnarray}
    i {\cal M}(\bar B_s \to \mu^+\mu^-) \Big|_{{\cal O}_7}^{\rm LO} &=&
    - \frac{\alpha}{2\pi} \, Q_\ell^2 \, Q_s \, C_7^{\rm eff} \, m \, M  f_{B_s} \, {\cal N} \, 
    \left[\bar u(p) \, (1+\gamma_5) \, v(p') \right] {\cal F}^{\rm LO}(E,m)
    \,,
\end{eqnarray}
with 
\begin{eqnarray}
    {\cal N} &=& V_{tb} \, V_{ts}^* \, \frac{4G_F}{\sqrt2} \, \frac{\alpha}{4\pi}
\end{eqnarray}
and 
\begin{eqnarray}
    {\cal F}^{\rm LO} (E,m) &=& \int_0^\infty \frac{d\omega}{\omega} \, \phi_+(\omega) 
    \left[ \frac12 \, \ln^2 \frac{m^2}{2E\omega} +  \ln \frac{m^2}{2E\omega} + \frac{\pi^2}{3} \right] \,.
    \label{eq:LO} 
\end{eqnarray}
Here $m$ denotes the muon mass, $M=m_{B_s} \simeq m_b$ the mass of the $\bar B_s$ meson, and $E\simeq M/2$ the muon energy in the $\bar B_s$ meson rest frame. 
The fermion electric charge fractions are given by $Q_\ell=-1$ and $Q_s =-1/3$.

In Ref.~\cite{Beneke:2017vpq} it has been pointed out that the strange-quark propagator in the box diagram leads to  a power-enhancement compared to the leading contribution from the local $b \to s\ell^+\ell^-$ operator
${\cal O}_{10}$. Furthermore, the propagator depends on the light-cone momentum fraction $\omega$ of the initial $\bar s$-quark in the $\bar B_s$ meson, and consequently the result involves a convolution with the hadronic light-cone distribution amplitude $\phi_+(\omega)$. Most importantly for our work, the form-factor ${\cal F}$ turns out to receive a double-logarithmic enhancement which can be traced back to endpoint divergences appearing in the convolution integrals from individual momentum regions, which we discuss in detail in the following.

\subsection{Kinematics}

We first fix the notation for the relevant kinematic variables.
The momenta of the muons in the final state
are used to define two light-cone vectors $n^\mu$ and $\bar n^\mu$, such that 
 \begin{eqnarray}
 \mbox{outgoing $\mu^-$ 
 } &: \quad & p^\mu = (\bar n\cdot  p) \, \frac{ n^\mu}{2} + 
 \frac{m^2}{(\bar n \cdot p)} \, \frac{\bar n^\mu}{2} \,, \nonumber \\[0.3em]
 \mbox{outgoing $\mu^+$ 
 }
 &:\quad & p'{}^\mu = (n \cdot p') \, \frac{\bar n^\mu}{2} + \frac{m^2}{(n \cdot p')} \, \frac{n^\mu}{2} \,,
 \end{eqnarray}
 and the momentum of the incoming $\bar B_s$-meson  given by
\begin{eqnarray}
    M v^\mu = M \left( \frac{n^\mu}{2} + \frac{\bar n^\mu}{2} \right) \,.
\end{eqnarray}
Here $n^2=\bar n^2=0$ and $n \cdot \bar n=2$. 
The energy $E$ of the muons in the $\bar B_s$-meson rest frame is approximately given by the relation
$(\bar n \cdot p)  = (n \cdot p') \simeq 2E = M$.
In the diagram in Fig.~\ref{fig:box}, the incoming $b$-quark is described as a static quark in heavy-quark effective theory,
characterized by a momentum $p_b^\mu = m_b  v^\mu  + \Delta^\mu$, with a residual momentum $\Delta^\mu$ of 
order $\Lambda_{\rm QCD}$. The momentum of the incoming $\bar s$-quark is decomposed as
\begin{eqnarray}
 \mbox{incoming $\bar s$-quark
 } &:\quad & l^\mu = (\bar n \cdot l) \, \frac{n^\mu}{2} + l_\perp^\mu + (n \cdot l) \, \frac{\bar n^\mu}{2} \,,
 \end{eqnarray}
where here and in the following we neglect the strange-quark mass.
It turns out that in the above diagram only the projection $\omega\equiv (\bar n \cdot l)$ of the $\bar s$-quark in the $\bar B_s$-meson  enters at leading approximation, which  therefore is identified as the argument of the light-cone distribution amplitude in Eq.~(\ref{eq:LO}). 
The ratio of $(\bar n\cdot p')$ and $(\bar n\cdot l)$ defines the relevant 
small expansion parameter which we denote as
	\begin{eqnarray}
		\lambda^2 = \frac{m^2}{2E\omega} \sim {\cal O}\left( \frac{\Lambda_{\rm QCD}}{m_b}\right) \sim {\cal O}\left( \frac{m}{m_b} \right) \,,
	\end{eqnarray}
where the muon mass $m$ and the light-cone projection $\omega$ of the light spectator-quark momentum are both counted to be of the order of the intrinsic QCD scale $\Lambda_{\rm QCD}$, which defines the \emph{soft} scale in the process. The energy of the muons, $ E \simeq M/2$, defines the \emph{hard} scale in the process. The product $(2E\omega)$ appears in the denominator of the strange-quark propagator (see below), and thus $\sqrt{2E\omega}$ defines the \emph{hard-collinear} scale in the process. 
    
\subsection{Momentum regions and different choices of analytic regulators}

We characterise the external and internal momenta by the scaling of the light-cone projections and the transverse components, using a short-hand notation, e.g.\
\begin{align}
		& p^\mu \;\,\, \longleftrightarrow \;  \frac{1}{2E} \, (\bar n\cdot p, p_\perp, n \cdot p) \sim (1,\lambda^2,\lambda^4) &\quad & \mbox{\small for collinear modes,} \\[0.8em] 
		& p'{}^\mu \; \longleftrightarrow \; \frac{1}{2E} \, (\bar n\cdot p', p'_\perp, n \cdot p')\sim (\lambda^4,\lambda^2,1) && \mbox{\small for anti-collinear modes,} \\[0.8em]
		& l^\mu \;\;\, \longleftrightarrow \; \frac{1}{2E} \, (\bar n\cdot l, l_\perp, n \cdot l)\sim (\lambda^2,\lambda^2,\lambda^2) && \mbox{\small for soft modes.} 
\end{align}
The presence of endpoint-divergent integrals requires the introduction of an additional regulator, since dimensional regularisation alone is not sufficient. Depending on the additional regulator, one then identifies different sets of relevant momentum regions.
In what follows, we compare two different choices for an analytic regulator,
\begin{eqnarray}
	\mbox{option (a)}  &:& \qquad 
	{{\cal R}_a(k)} = \left(\frac{\nu^2}{-(n \cdot k)(\bar n\cdot l) + i 0} \right)^\delta \,, 
	\\
	\mbox{option (b)} &:& \qquad 	{{\cal R}_b(k)} =	\left( \frac{\nu^2}{(\bar n \cdot k)(n \cdot p') - (n\cdot k) (\bar n \cdot l) + i 0}\right)^\delta \, ,
\end{eqnarray}
which are assigned to a muon propagator with momentum $k$. These regulators are a generalization of the corresponding options discussed -- for instance -- in Ref.~\cite{Bell:2022ott} (see also references therein) written
in a manifestly boost-invariant form, using the external momenta 
$(\bar n \cdot l)$ and $(n \cdot p')$.

The non-vanishing momentum regions for the box diagram in Fig.~\ref{fig:box} are summarised in Table~\ref{tab:regions}.  Here the momenta of the upper and lower photon lines are given by
    \begin{eqnarray}
    	q^\mu = (p-k)^\mu \,, \qquad  q'{}^\mu = (p'+k)^\mu \,,
    \end{eqnarray}
    and the intermediate strange-quark propagator has momentum
    \begin{eqnarray}
    	k'{}^\mu = (p'+k-l)^\mu  \,.
    \end{eqnarray}

\begin{table}[t]
\begin{center}
    \begin{tabular}{c||c|c|c|c||c}
    \hline 
    region & muon  & upper photon & $\bar s$-quark & lower photon & regulator 
    \\
    \hline \hline 
 $\overline{hc}$  
 & $k \sim (\lambda^2,\lambda,1)$ 
 & $q \sim (1,\lambda,1)$
 & $k' \sim (\lambda^2,\lambda,1)$
 & $q' \sim (\lambda^2,\lambda,1)$
 & ${\cal R}_{a,b}$
 \\
 \hline 
 $\overline{c}$  
 & $k \sim (\lambda^4,\lambda^2,1)$ 
 & $q \sim (1,\lambda^2,1)$
 & $k' \sim (\lambda^2,\lambda^2,1)$
 & $q' \sim (\lambda^4,\lambda^2,1)$
 & ${\cal R}_{a,b}$
 \\
 \hline 
 $s$  
 & $k \sim (\lambda^2,\lambda^2,\lambda^2)$ 
 & $q \sim (1,\lambda^2,\lambda^2)$
 & $k' \sim (\lambda^2,\lambda^2,1)$
 & $q' \sim (\lambda^2,\lambda^2,1)$
 & ${\cal R}_{a,b}$
 \\
 \hline 
 $\overline{sc}$  
 & $k \sim (\lambda^3,\lambda^2,\lambda)$ 
 & $q \sim (1,\lambda^2,\lambda)$
 & $k' \sim (\lambda^2,\lambda^2,1)$
 & $q' \sim (\lambda^3,\lambda^2,1)$
 & ${\cal R}_{b}$
 \\
 \hline 
 \hline 
    \end{tabular}
\end{center}
\caption{\label{tab:regions} Momentum regions contributing to the box diagram in Fig.~\ref{fig:box}. 
Here, $\overline{hc}$, $\bar c$, $s$, and $\overline{sc}$ denote the anti-hard collinear,   anti-collinear, soft, and anti-soft-collinear regions, respectively.
Notice that the anti-soft-collinear region yields scaleless integrals if the analytic regulator ${\cal R}_a$ is used.}
\end{table}

We observe that in all relevant regions the intermediate strange-quark  has anti-hard-collinear virtuality,
$(k')^2 \sim \lambda^2$, with 
$$
  k' \sim \left\{ \begin{array}{lcl} (\lambda^2,\lambda,1) && \mbox{for $k$ anti-hard-collinear,} \\
  	(\lambda^2,\lambda^2,1) && \mbox{for $k$ soft, anti-collinear, or anti-soft-collinear.}
\end{array} \right. $$
This implies that in all relevant cases, the strange-quark propagator
can effectively be replaced by 
\begin{eqnarray} 
  S_F(k') &\simeq& i \, \frac{\slashed {\bar n}}{2} \, 
  \frac{(n \cdot k')}{(n \cdot k')(\bar n \cdot k') + (k_\perp^\prime)^2+i0}
  \nonumber \\[0.25em]
  &\simeq& i \, \frac{\slashed {\bar n}}{2} \, \frac{(2E + n \cdot k)}{(2E +n \cdot k)(\bar n \cdot k- \omega) + k_\perp^2+i0}
  \,,
\end{eqnarray}
which further simplifies in the individual momentum regions (see below). Here $(n \cdot k') \sim 2E$ 
is the large component of the anti-hard-collinear momentum, and the propagator only depends on the light-cone projection $\omega=\bar n\cdot l$ of the strange-quark momentum in the $\bar B_s$ meson.

Inserting this approximation into the box diagram and performing the Lorentz contractions and Dirac projections, and exploiting the equations of motion for the external muon states, we find that the problem reduces to analysing the scalar integral 
\begin{equation}
\begin{aligned}
	I(\omega) &= \int \! d (\bar n \cdot k) \, \int \! d (n \cdot k) \, 
	\int \widetilde{dk_\perp} \,  \cdot \frac{1}{(p-k)^2+i0}
        \cdot\frac{1}{(p'+k)^2+i0} 
	\cr &\times
	 \frac{(2E + n \cdot k) \, 2E\omega}{(2E+n \cdot k)(\bar n \cdot k-\omega) + k_\perp^2+i0}
	\cdot \frac{2E + n \cdot k}{(n \cdot k)(\bar n \cdot k) + k_\perp^2 -m^2 +i0} 
	\, \Bigg|_{\mbox{\scriptsize leading power}} .
\label{Iomega}
\end{aligned}
\end{equation}
The integral measure for the transverse loop momentum reads
$$\displaystyle \widetilde {dk_\perp} =
\frac{i \mu^{2\epsilon} \,  e^{\epsilon \gamma_E}}{2\pi^{D/2}} \, d^{D-2}k_\perp \, ,$$
such that 
\begin{equation}
 {\cal F}^{\rm LO}(E,m) = 
 \int_0^\infty \frac{d\omega}{\omega} \, \phi_+(\omega) \, I(\omega) \, ,
 \label{FFandIw}
\end{equation}
where $\gamma_E$ is the Euler-Mascheroni constant, $D=4-2\epsilon$ denotes the number of space-time dimensions, and $\mu$ refers to the $\overline{\rm MS}$ scale.
We now discuss the contributions to the integral $I(\omega)$ from the different momentum regions in turn.

\subsubsection{The anti-hard-collinear region}
 The anti-hard-collinear region in the integral (\ref{Iomega}) is defined by the scaling
$k  \sim  (\lambda^2,\lambda,1)$, which simplifies the integrand at leading power to
\begin{equation}
\begin{aligned}
	I_{\overline{hc}}(\omega) &= \int d(\bar n \cdot k) \, \int d (n \cdot k) \, 
	\int \widetilde{dk_\perp} \, \\[0.3em]
	&\times \, \frac{1}{- 2E \, (n \cdot k)+i0}
	\cdot\frac{1}{(2E+n \cdot k)(\bar n \cdot k) + k_\perp^2+i0} \\[0.3em]
	&\times \, 
	\frac{(2E + n \cdot k) \, 2E\omega}{(2E+n \cdot k)( \bar n \cdot k- \omega) + k_\perp^2+i0}
	\cdot \frac{2E + n \cdot k}{(n \cdot k)(\bar n \cdot k) + k_\perp^2 +i0} \, .
\end{aligned}
\end{equation}
This integral features an endpoint divergence for $(n \cdot k) \to 0$, which is regularized in $D=\nobreak 4-\nobreak 2\epsilon$ dimensions, such that the analytic regulator can be dropped.
Actually, the endpoint divergence arises from the convolution of a hard eikonal propagator (the upper photon)
with a remaining jet function
consisting of three anti-hard-collinear propagators.  

It is convenient to
 first perform the $(\bar n \cdot k)$ integration by residues, 
and then afterwards the $k_\perp$ integral.
This can be written as
\begin{eqnarray}
	I_{\overline{hc}}(\omega) &=&
    \int_0^1 \frac{du}{u} \,  H_1^{(0)}(u) \,
	\bar J_1^{(1)}(u;\omega)
	\,,
\end{eqnarray}
where $
	H_1^{(0)}(u) = 1 
$
is the leading-order value of the hard matching coefficient for the 
$b \to s\gamma$ tensor current with an energy transfer $(1-u) E$,
and
\begin{eqnarray}
	 \bar J_1^{(1)}(u;\omega) 
	 = -  \Gamma(\epsilon)  \left( \frac{\mu^2 e^{\gamma_E}}{2E\omega u(1-u)} \right)^\epsilon (1-u)
	 \,,
	 \label{H1J1}
\end{eqnarray}
is the first non-vanishing term in the jet function associated to the anti-hard-collinear momentum region,
with $u\equiv -(n \cdot k)/(n \cdot p')$.
Performing the $u$-integral and expanding in $\epsilon$, the result for the anti-hard-collinear region gives
\begin{eqnarray}
	I_{\overline{hc}}(\omega) = 
		\frac{1}{\epsilon^2} + \frac1{\epsilon} \, \ln \frac{\mu^2}{2E\omega} + \frac12 \, \ln^2\frac{\mu^2}{2E\omega} -  \frac{\pi^2}{12}
		+ \frac{1}{\epsilon} + \ln \frac{\mu^2}{2E\omega} + 2 \, ,
\end{eqnarray}
which features the standard double-logarithms from soft and collinear infrared (IR) divergences, and a single logarithm arising from an endpoint-finite longitudinal integration.

\subsubsection{The anti-collinear region}

The anti-collinear region in the integral (\ref{Iomega}) is defined by the scaling
$
  k \sim (\lambda^4,\lambda^2,1)
$, which simplifies the integrand at leading power to
\begin{equation}
\begin{aligned}
	I_{\bar c}(\omega) &= \int \! d (\bar n\cdot k) \, \int \! d (n \cdot k) \, 
	\int \widetilde{dk_\perp} \,  \\[0.3em]
	&\times \, \frac{1}{-2E\, (n \cdot k)+i0}
	\cdot\frac{1}{(2E+n \cdot k)(\bar n \cdot p' +\bar n \cdot k)+ k_\perp^2+i0}  \\[0.3em] 
	&\times \,
	\frac{2E\omega}{-\omega +i0}
	\cdot \frac{2E + n \cdot k}{(n \cdot k)(\bar n \cdot k) + k_\perp^2 - m^2 +i0} \left( \frac{\nu^2}{- \omega \, (n \cdot k) + i0}\right)^\delta \, ,
\end{aligned}
\end{equation}
with $(\bar n \cdot p')=m^2/2E$.
This integral now features an endpoint singularity from $(n \cdot k) \to 0$ which is \emph{not} regularized in $D\neq 4$ dimensions due to the non-vanishing mass in the anti-collinear muon propagator.
In this case the analytic regulator has been kept, and because of the anti-collinear scaling of $k$ both variants ${\cal R}_a$ and ${\cal R}_b$ reduce to the same term as indicated above.

We note that the integral now involves two eikonal propagators. The first one, $(-(\bar n\cdot p)(n\cdot k))^{-1}$,
reflects the same tree-level hard function from the upper photon propagator as in the anti-hard-collinear region. The second one, $(-\omega)^{-1}$, reflects the tree-level anti-hard-collinear function which decouples from the anti-collinear loop integral.
Again, we can first perform the $(\bar n \cdot k)$ integration by residues, and then the $k_\perp$ integral in $D-2$ dimensions. The result can be written as
\begin{eqnarray}
	I_{\overline{c}}(\omega) &=&
    \bar J_{2}^{(0)}(1,\omega) \, \int_0^1 \frac{du}{u} \,  H_1^{(0)}(u) \,
	\bar C^{(1)}(u;\omega) \,,
\end{eqnarray}
where we have defined the leading-order term in the jet function 
for the anti-hard-collinear strange-quark propagator with $(\bar n\cdot k') = z \, \omega$, 
\begin{eqnarray} 
\bar J_2^{(0)}(z,\omega) &=& \frac{1}{z} \label{barJ2} \,,
\end{eqnarray}
where the overall factor $1/\omega$ --~that appears in the convolution with the $B_s$-meson LCDA~--
has been factored out.
The function  
\begin{eqnarray}
   \bar C^{(1)}(u;\omega)  
    &=& 
	 \Gamma(\epsilon) \,
\left(\frac{\mu^2 e^{\gamma_E}}{m^2}\right)^{\epsilon}
		\left( \frac{\nu^2}{2E\omega} \right)^\delta
	\left( 1-u \right)^{1-2\epsilon} \, u^{-\delta} \,,
	\label{C1}
\end{eqnarray}
refers to the leading term of the anti-collinear function that 
describes the recombination process $\mu^+(-k) \, \gamma(p'+k) \to \mu^+(p')$.
Performing the $u$-integration, the final result for the anti-collinear region gives,
\begin{eqnarray}
	I_{\overline{c}}(\omega) &=& 
	\left( -\frac{1}{\delta} - \ln \frac{\nu^2}{2E\omega} \right) 
	\left( \frac{1}{\epsilon} + \ln \frac{\mu^2}{m^2} \right)
	+ \frac{\pi^2}{3} 
	- \frac{1}{\epsilon} - \ln \frac{\mu^2}{m^2} -2 \, ,
\end{eqnarray}
where we first had to expand in $\delta$ and subsequently in $\epsilon$.
We observe that the anti-collinear region depends on the soft momentum $\omega$ 
 through the analytic regulator, reflecting the collinear-anomaly phenomenon \cite{Becher:2010tm,Becher:2011pf}.

\subsubsection{The soft region}

The soft region in the integral (\ref{Iomega}) is defined by the scaling
$
  k \sim (\lambda^2,\lambda^2,\lambda^2) 
$, which simplifies the integrand at leading power to
\begin{equation}
\begin{aligned}
	I_s(\omega) &= \int d (\bar n \cdot k)
\, \int d (n \cdot k) \, 
\int \widetilde{dk_\perp} \, 
 \cdot \frac{1}{-2E \, (n \cdot k)  +i0}
\cdot\frac{1}{2E \, (\bar n \cdot k)+i0}
\cr 
&\times
\frac{2E \omega}{\bar n \cdot k- \omega + i0}
\cdot \frac{2E}{(n \cdot k)(\bar n \cdot k) + k_\perp^2 -m^2 +i0} \,
{\cal R}_{a,b}(k)
\,,
\end{aligned}
\end{equation}
which is again endpoint-divergent and thus requires the additional regulator, as indicated.
Let us first consider the analytic regulator for case (b) which, in the soft region, reduces to 
$$
{\cal R}_{b}(k) \to \left( \frac{\nu^2}{2 E \, (\bar n \cdot k) + i0
} \right)^\delta \,,
$$
where the $i0$ prescription matches the expression in the respective eikonal photon propagator.
In that case, one can first perform the $(n \cdot k)$ integration by residues, and then the $k_\perp$ integral in $D-2$ dimensions.
The soft contribution to the integral then takes the form 
\begin{eqnarray}
(b): \qquad 
I_{s}(\omega) & = & 
	\Gamma(\epsilon) \, 
	\left(\frac{\mu^2 e^{\gamma_E}}{m^2}\right)^{\epsilon}
	\left( \frac{\nu^2}{2E\omega}\right)^\delta
	\int\limits_0^{\infty} dv \, \frac{(v+i0)^{-1-\delta}}{v-1 + i 0} 
	\,,
\end{eqnarray}
with $v\equiv (\bar n \cdot k)/\omega$.
The endpoint divergence at $v \to 0$ is regularized by the analytic regulator, while the integral converges in the ultraviolet (UV), $v \to \infty$.
Notice that the integral generates an imaginary part which stems from the situation when the strange-quark propagator goes on-shell for $v \to 1$, 
\begin{eqnarray}
(b) \qquad	I_{s}(\omega) &=& 
	\left( \frac{1}{\epsilon} + \ln \frac{\mu^2}{m^2} \right) \left(   -\frac{1}{\delta} - \ln \frac{\nu^2}{2E\omega} +i\pi  \right) 
	\,.
\end{eqnarray}

On the other hand, taking the analytic regulator for case (a)
$$
 {\cal R}_a(k) \to \left( \frac{\nu^2}{-\omega \, (n \cdot k) + i0} \right)^\delta \,,
$$
we can first perform the $k_\perp$ integration, to end up with
\begin{equation}
\begin{aligned}
\mbox{(a)} \qquad \tilde I_s(\omega) &= 
    \frac{\Gamma(\epsilon)  }{2\pi i}
	\, \int \frac{d(n \cdot k)}{-n \cdot k + i 0} \, 
	\int\frac{d(\bar n \cdot k)}{\bar n \cdot k + i0} \, \frac{\omega}{\bar n\cdot k - \omega + i0}
	\\* &\times
	\left( \frac{\mu^2 e^{\gamma_E}}{-(n \cdot k)(\bar n \cdot k)+m^2-i0} \right)^\epsilon \, {\cal R}_a(k) \,.
\end{aligned}
\end{equation}
Now consider the analytic structure in $(\bar n \cdot k)$-plane:
\begin{itemize}
	\item There are two poles at 
	\begin{eqnarray*}
	{\rm Re} (\bar n \cdot k) = 0,\, \omega  \quad \mbox{\emph{below} the real axis.}
	\end{eqnarray*}
	\item There is a branch cut from the dimensional regulator,  
	\begin{eqnarray*}
		\mbox{for $( n \cdot k)>0$} \quad &:\quad &  m^2/(n \cdot k)
		\leq {\rm Re}(\bar n \cdot k), \quad \mbox{\emph{below} the real axis,}
		\cr 
		\mbox{for $( n \cdot k)<0$} \quad &:\quad&  {\rm Re}(\bar n \cdot k) \leq  m^2/(n \cdot k), \quad \mbox{\emph{above} the real axis.}
	\end{eqnarray*} 
     \end{itemize} 
  For $(n \cdot k) > 0$, the integrand is analytic in the upper half plane, and therefore the integral vanishes. For $(n \cdot k) <0$, the branch cut from the dimensional regulator is moved to the upper half plane, and by picking up the corresponding discontinuity
    \begin{align}
      {\rm Disc} \left( \frac{\mu^2}{-(n \cdot k)(\bar n \cdot k)+m^2-i0} \right)^\epsilon = \frac{2\pi i \, \theta((n \cdot k)(\bar n \cdot k)-m^2)}{\Gamma(\epsilon)\Gamma(1-\epsilon)}\left( \frac{\mu^2}{(n \cdot k)(\bar n \cdot k)-m^2} \right)^\epsilon \,,
      \label{eq:disc}
      \end{align}
we arrive at the representation 
\begin{equation}
\begin{aligned}
\mbox{(a)} \qquad 
\tilde I_s(\omega) &=
  \frac{1}{\Gamma(1-\epsilon)} \, \int\limits_{-\infty}^0 \frac{d(n \cdot k)}{n \cdot k} \, 
	\int\limits_{-\infty}^{m^2/n\cdot k}\frac{d(\bar n \cdot k)}{\bar n \cdot k} \, \frac{\omega}{ 
	\omega - \bar n\cdot k }
	\left( \frac{\mu^2 e^{\gamma_E}}{(n \cdot k)(\bar n \cdot k)-m^2} \right)^\epsilon \, {\cal R}_a(k)
	\cr 
	&=   \frac{1}{\Gamma(1-\epsilon)} \left( \frac{\mu^2  e^{\gamma_E}}{2E\omega} \right)^\epsilon \left(\frac{\nu^2}{2E\omega} \right)^\delta
 \int_0^\infty \frac{du}{u} \, 
	\int_{\lambda^2/u}^\infty \frac{d\rho}{\rho} \, \frac{1}{1+\rho}
	\left(u\rho-\lambda^2\right)^{-\epsilon} \, u^{-\delta} \, ,
\end{aligned}
\end{equation}
where $\rho = - (\bar n \cdot  k)/\omega$.
This can be written as 
\begin{equation}
\mbox{(a)} \qquad \tilde I_s(\omega) = 
 H_1^{(0)}(0) \, \int_0^\infty \frac{du}{u} \, \int_0^\infty \frac{d\rho}{\rho} \, S^{(1)}(u,\rho;\omega) \, \bar J_2^{(0)}(1+\rho,\omega)
\end{equation}
with the same hard and jet function as already defined above, and a soft function
\begin{equation}
    S^{(1)}(u,\rho;\omega) = \theta(u\rho -\lambda^2) \left( \frac{\mu^2  e^{\gamma_E}}{2E\omega} \right)^\epsilon \left( \frac{\nu^2}{2uE\omega} \right)^\delta   \frac{(u\rho -\lambda^2)^{-\epsilon}}{ \Gamma(1-\epsilon)} \, \,,
    \label{S1}
\end{equation}
which stems from the discontinuity of the muon propagator.
Performing the longitudinal integrations, we end up with 
\begin{eqnarray}
(a) \qquad     	\tilde I_s(\omega) &=&
\left( \frac{1}{\delta} + \ln \frac{\nu^2}{m^2}\right)\left( \frac{1}{\epsilon} + \ln \frac{\mu^2}{m^2}\right) - \frac{1}{\epsilon^2} - \frac{1}{\epsilon} \, \ln \frac{\mu^2}{m^2}  - \frac12 \, \ln^2 \frac{\mu^2}{m^2} + \frac{\pi^2}{12}
\,.
\end{eqnarray}
In that case, the full integral is already reproduced by the sum of the anti-hard-collinear, the anti-collinear, and the soft region,
\begin{eqnarray}
(a) \qquad 
I(\omega) &=& I_{\overline{hc}}(\omega) + I_{\overline c}(\omega) + \tilde I_s(\omega) 
	= \frac12 \, \ln^2 \frac{m^2}{2E\omega} 
	+ \ln \frac{m^2}{2E\omega} + \frac{\pi^2}{3} \,.
\end{eqnarray}
Notice that the single logarithm in this expression stems from the cancellation of single powers of $1/\epsilon$ in the $\overline{hc}$ and $\overline c$ region.

\subsubsection{Anti-soft-collinear region}
For the regulator ${\cal R}_b$, we also have to take into account an anti-soft-collinear region, where
$
 k^\mu \sim (\lambda^3,\lambda^2,\lambda) \,.
$
In this region the integral (\ref{Iomega}) at leading power reduces to
\begin{equation}
\begin{aligned}
	I_{\overline{sc}}(\omega) &= \int \! d (\bar n \cdot k) \, \int \! d (n \cdot k) \, 
	\int \widetilde{dk_\perp} \,  \cdot \frac{1}{-2E \, n \cdot k+i0}
        \cdot\frac{1}{2E \, \bar n \cdot k+i0} 
	\cr &\times
	 \frac{4E^2\omega}{-2E \omega +i0}
	\cdot \frac{2E}{(n \cdot k)(\bar n \cdot k) + k_\perp^2 -m^2 +i0}  \, {\cal R}_{ b}(k)
	\,.
\end{aligned}
\end{equation}
In that case, it is again convenient to first perform the $k_\perp$ integration, which yields,
\begin{equation}
\begin{aligned}
(b)\quad 	I_{\overline{sc}}(\omega) &= 
    \frac{\Gamma(\epsilon) }{2\pi i}
	\, \int \frac{d(n \cdot k)}{n \cdot k - i 0} \, 
	\int\frac{d(\bar n \cdot k)}{\bar n \cdot k + i0} \, 
	\\* &\times
	\left( \frac{\mu^2  e^{\gamma_E}}{-(n \cdot k)(\bar n \cdot k)+m^2-i0} \right)^\epsilon 
	\left( \frac{\nu^2}{2E \, (\bar n \cdot k) - \omega \, (n \cdot k) + i0} \right)^\delta \,.
\end{aligned}
\end{equation}
We now consider the analytic structure of the integrand in the complex $(n \cdot k)$ plane.
\begin{itemize}
	\item There is always a pole at 
	\begin{eqnarray*}
	{\rm Re} (n \cdot k) = 0,  \quad \mbox{\emph{above} the real axis.}
	\end{eqnarray*}
	\item There is a branch cut from the dimensional regulator,  
	\begin{eqnarray*}
		\mbox{for $(\bar n \cdot k)>0$} \quad &:\quad &  m^2/(\bar n \cdot k)
		\leq {\rm Re}(n \cdot k), \quad \mbox{\emph{below} the real axis,}
		\cr 
		\mbox{for $(\bar n \cdot k)<0$} \quad &:\quad&  {\rm Re}(n \cdot k) \leq  m^2/(\bar n \cdot k), \quad \mbox{\emph{above} the real axis.}
	\end{eqnarray*}  
    \item There is a branch cut from the rapidity regulator,
     \begin{eqnarray*}
     	(n \cdot p')(\bar n \cdot k)/\omega \leq {\rm Re}(n \cdot k) \,, \quad  \mbox{\emph{above} the real axis.}
     \end{eqnarray*} 
     \end{itemize} 
  For $(\bar n \cdot k) < 0$, the integrand is analytic in the lower half plane, and therefore the integral vanishes. For $(\bar n \cdot k) >0$, the branch cut from the dimensional regulator is moved to the lower half plane, and by picking up the corresponding discontinuity (\ref{eq:disc}),
we obtain
   \begin{eqnarray}
 	(b) \quad 
 	I_{\overline{sc}}(\omega) &=& 
 	\frac{1}{\Gamma(1-\epsilon)} \,
 	 \int\limits_0^\infty \frac{d(n\cdot k)}{n \cdot k} \, 
 	\int\limits_0^\infty \frac{d (\bar n \cdot k)}{\bar n \cdot k} \, \theta((n \cdot k)(\bar n \cdot k)-m^2)
 	\\* &\times& 
 	\left( \frac{\mu^2  e^{\gamma_E}}{(n \cdot k)(\bar n \cdot k)-m^2} \right)^\epsilon 
 	\left( \frac{\nu^2}{2E \, (\bar n \cdot k) - \omega \, (n \cdot k) + i0} \right)^\delta
 	\nonumber \\[0.3em] &=&
    \left( \frac{\mu^2  e^{\gamma_E}}{2E\omega}\right)^\epsilon  
 	\left( \frac{\nu^2}{2E\omega} \right)^\delta 
 	\, \int\limits_0^\infty \frac{du'}{u'} \, 
 	\int\limits_0^\infty \frac{dv}{v} \, 
 	\theta(u'v-\lambda^2) 
 	\frac{( u' v -\lambda^2)^{-\epsilon}}{ \Gamma(1-\epsilon)} 
 	\left( v-u'  + i0 \right)^{-\delta}
 	\,,
 	\cr &&
 \end{eqnarray}
 where $u'=-u=(n\cdot k)/2E$ and $v = (\bar n \cdot k)/\omega$ as before.
This integral can most easily be solved by decomposing it into two terms, one with $v >u^\prime$ and one with $v< u^\prime$. Both integrals are identical, except for an additional phase factor $e^{-i\pi\delta}$ that arises in the second case  and effectively amounts to replacing $1/\delta \to 1/\delta - i\pi$ after expanding in $\delta$. It is also instructive to consider the variable transform $x= v-u^\prime$ and $y= u^\prime v$. Here, the integral over $x$ produces a UV pole that is regularized by $\delta$, while $y$ produces a UV pole that is regularized by $\epsilon$. The net result is
\begin{equation}
\begin{aligned}
\!\!(b) \quad\!	I_{\overline{sc}}(\omega) &= \tilde I_s(\omega)-I_s(\omega) 
	\cr &=
	 - \frac{1}{\epsilon^2} - \frac{1}{\epsilon}  \ln \frac{\mu^2}{m^2} - \frac12  \ln^2\!\frac{\mu^2}{m^2}   + 
	 \!\left(\frac{1}{\epsilon} + \ln\!\frac{\mu^2}{m^2}\right)
	 \!\!\left(\frac{2}{\delta} -i\pi + \ln\! \frac{\nu^2}{2E\omega} + \ln\! \frac{\nu^2}{m^2}  \right)  
	 +  \frac{\pi^2}{12} \,.
\end{aligned}
\end{equation}

With this we obtain an alternative decomposition of the form-factor integral
\begin{eqnarray}
(b) \quad 	 I(\omega) &=& I_{\overline{hc}}(\omega) + I_{\overline c}(\omega) + I_s(\omega) + I_{\overline{sc}}(\omega) \, .
\end{eqnarray} 
Notice that the (divergent) imaginary part $i\pi \left(\frac{1}{\epsilon} + \ln \frac{\mu^2}{m^2} \right)$ cancels between the soft and anti-soft-collinear region. Furthermore, in the $\overline{sc}$ region, the reference scale for the analytic regulator is set by the geometric mean of the anti-hard-collinear scale $\sqrt{2E\omega}$ and the soft scale $m$.

\subsection{Anti-soft-collinear region with cut-off}

The double-logarithmic term in $I(\omega)$ can also be obtained in an alternative manner by considering the IR contribution to the anti-soft-collinear region only (see also the analogous discussion in Ref.~\cite{Bell:2022ott} or Ref.~\cite{Liu:2018czl} and references therein), where the dimensionless integration variables $u,v$ are cut off from above at their natural value $1$,
\begin{eqnarray}
I(\omega)\Big|_{\rm double-log}	= I_{\overline{sc}} \Big|_{\rm cut} &=& 
\int\limits_0^1 \frac{du}{u} \, 
	\int\limits_0^1 \frac{dv}{v} \, 
	\theta(uv-\lambda^2) = \frac12\, \theta(1-\lambda^2) \, \ln^2\lambda^2 
	\,. \label{scbarcut}
\end{eqnarray}
Notice that we have kept a theta-function $\theta(1-\lambda^2)$ here, because $\lambda^2=\frac{m^2}{2E\omega}$ depends on the integration variable $\omega$, such that \emph{formally} we can have $\lambda^2>1$ for $\omega < m^2/2E$. However, in the end, the hadronic parameters in the $B_s$-meson LCDA respect the 
power-counting $\omega \sim \Lambda_{\rm QCD}$, such that we can expand,
$\theta(1-\lambda^2) = 1 + {\cal O}(m/E)$.
The double-logarithmic term in the form factor for the ${\cal O}_7$ contribution can thus be written as
\begin{eqnarray}
	{\cal F}^{\rm LO}(E,m)\Big|_{\rm double-log} &=& 
	\int \frac{d\omega}{\omega} \, \phi_+(\omega) \, \int\limits_0^1 \frac{du}{u} \, 
	\int\limits_0^1 \frac{dv}{v} \, 
	\theta(2 E  \omega  \, uv-m^2)
	\,.
\end{eqnarray}
For the following discussion it is useful to rewrite the integral over the light-quark momentum $\omega$ in the so-called dual space introduced in Ref.~\cite{Bell:2013tfa}:
\begin{eqnarray}
	\phi_+(\omega) &=& \int \frac{d\omega'}{\omega'} \, \sqrt{\frac{\omega}{\omega'}} \, J_1\left( 2 \sqrt{\frac{\omega}{\omega'}}\right) \, \rho_+(\omega')
	\, ,
\end{eqnarray}
with the Bessel function of the first kind $J_1$. Since the second logarithmic moments of $\phi_+(\omega)$ and $\rho_+(\omega')$ coincide~\cite{Bell:2013tfa}, we can as well write the
double-logarithmic contribution to the form factor as 
\begin{eqnarray}
	{\cal F}^{\rm LO}(E,m)\Big|_{\rm double-log} &=& 
	\frac12 \, \int \frac{d\omega'}{\omega'} \, \rho_+(\omega') \, 
	\ln^2 \frac{m^2 e^{2\gamma_E}}{2E \omega'} 
	\,.
	\label{Fdoublelog}
\end{eqnarray}

\subsection{Construction of the factorization theorem}

The analysis of the box diagram above determines the form of a bare factorization theorem that takes into account additional QCD corrections from different momentum regions, depending on the regulator being used. For concreteness, we stick to the analytic regulator ${\cal R}_a(k)$, such that the anti-soft-collinear region do not appear in the bare factorization theorem. The form factor for the ${\cal O}_7$ contribution then decomposes into three additive terms,
\begin{eqnarray}
{\cal F}(E,m) &=& \int_0^\infty \frac{d\omega}{\omega} \, \phi_+(\omega) 
\Bigg\{ \int_0^1 \frac{du}{u} \, H_1(u) \, \bar J_1(u;\omega) \,
\cr 
&& \qquad {} + \bar J_2(1,\omega) \, \int_0^1 \frac{du}{u} \, H_1(u) \, \bar C(u;\omega) 
\cr 
&&
\qquad {} + H_1(0) \, \int_0^\infty \frac{du}{u} \, \int_0^\infty \frac{d\rho}{\rho} \, S(u,\rho;\omega) \, \bar J_2(1+\rho,\omega)  \Bigg\}_{\rm bare}
\label{eq:bare}
\end{eqnarray}
where each individual term contains an endpoint-divergent convolution integral that is regularised for finite $\epsilon$ and $\delta$.
Here  
\begin{align}
H_1(u) & = H_1^{(0)}(u) + {\cal O}(\alpha_s) \, , \\
\bar J_1(u;\omega) & = \bar J_1^{(1)}(u;\omega) + {\cal O}(\alpha_s) \, , \\
\bar J_2(z) & = \bar J_2^{(0)}(z) + {\cal O}(\alpha_s) \,,
\end{align}
with $H_1^{(0)}(u)=1$, and $\bar J_1^{(1)}(u;\omega)$, $\bar J_2^{(0)}(z)$ given in 
Eqs.~(\ref{H1J1}) and~(\ref{barJ2}).
Notice that in the bare factorization theorem (\ref{eq:bare}) we have not taken into account 
additional QED corrections to the $\gamma^*\gamma^* \to \mu^+\mu^-$ subprocess, i.e. the collinear and soft functions related to the muon,
\begin{eqnarray}
   \bar C(u;\omega) &=& \bar C^{(1)}(u;\omega)
+ {\cal O}(\alpha) \,,
	\label{eq:Cbar1} \\
S(u,\rho;\omega) &=& S^{(1)}(u,\rho;\omega) 
+ {\cal O}(\alpha)
\label{eq:S1}
\end{eqnarray}
are set to their leading-order expressions given in Eqs.~(\ref{C1}) and~(\ref{S1}).

In order to get rid of the analytic regulator in the bare factorisation theorem, we have to exploit the refactorisation condition 
\cite{Boer:2018mgl,Liu:2018czl,Bell:2013tfa} 
for the anti-collinear function.
Following the notation of~\cite{Liu:2018czl}, we define
\begin{eqnarray}
&& \left[\left[\bar C(u;\omega)\right]\right] \equiv \bar C(u;\omega)\big|_{u\to 0}
= \int_0^\infty \frac{d\rho}{\rho} \, S(u,\rho;\omega) + {\cal O}(\alpha)
\label{refac}
\end{eqnarray}
which in our case provides a trivial identity between the QED fixed-order expressions in (\ref{C1}) and (\ref{S1}).
In addition, we use 
that in case of analytic regulators the scaleless double-integral
\begin{eqnarray}
\int_0^\infty \frac{du}{u} \, \int_0^\infty \frac{d\rho}{\rho} \, S(u,\rho;\omega) &=& 0 \,,
\end{eqnarray}
vanishes.
As described in detail in the appendix~\ref{app}, the factorization can then be re-arranged as follows,
\begin{eqnarray}
{\cal F}(E,m) &=& \int_0^\infty \frac{d\omega}{\omega} \, \phi_+(\omega) 
\Bigg\{ \int_0^\infty \frac{du}{u} \Big[ H_1(u) \, \bar J_1(u;\omega) \, \theta(1-u) 
\cr && \qquad \qquad \qquad \qquad   {}
- H_1(0) \, \bar J_2(1,\omega) \, \theta(u-1) \, \int_1^\infty\frac{d\rho}{\rho} \, S(u\rho;\omega) \Big]_{\lambda^2 \to 0} 
\cr 
&& \qquad {} + \bar J_2(1,\omega) \, \int_0^1 \frac{du}{u} \Big[ H_1(u) \,  \bar C(u) - H_1(0) \left[\left[ \bar C(u)\right]\right] \Big]
\nonumber\\[0.5em]
&& \qquad {} +H_1(0) \, \int_0^\infty \frac{du}{u} \, \int_0^\infty \frac{d\rho}{\rho} \left[ \bar J_2(1+\rho,\omega) -  
\theta(1-\rho) \, \bar J_2(1,\omega) \right] S(u \rho;\omega)
\nonumber\\[0.3em]
&&  \qquad {} + \bar J_2(1,\omega) \, H_1(0) \, \int_0^1 \frac{du}{u}\, \int_0^1 \frac{d\rho}{\rho} \,  S(u \rho;\omega) \Big|_{\rm \lambda^2\to 0} \Bigg\}
\label{fact}
\end{eqnarray}
where now the limit $\delta \to 0$ can be performed \emph{before} the integrations, and therefore the anti-collinear function does not depend on $\omega$ anymore, and the soft function only depends on the product $u\rho$ (and, via the dimensional regulator, on $\omega$).
Comparison with the bare factorization theorem shows that now all endpoint divergences are subtracted from the anti-hard-collinear, anti-collinear and soft momentum regions.
The endpoint logarithms are fully contained in the fourth term which coincides with the expression obtained from the anti-soft-collinear region with explicit momentum cut-offs, see Eq.~(\ref{scbarcut}) above. Notice that the emerging cut-offs in the first and fourth line lead to additional power corrections in $\lambda^2$ which should be dropped, as indicated by the limit $\lambda^2 \to 0$.
The factorization theorem in Eq.~(\ref{fact}) represents one of the main results of our paper. The various functions appearing in Eq.~(\ref{fact}) can now be associated to effective operators in soft-collinear effective theory (SCET) which can be renormalized in a standard manner \emph{before} the convolution integrals are performed. However, a detailed proof of the factorization theorem along these lines is beyond the scope of this work.
We should also stress that despite the fact that the structure of our factorisation theorem allows us to systematically include QCD corrections, it does not capture the effect of additional QED corrections. In particular, the latter would also modify the refactorisation condition~(\ref{refac}), and the QED corrections to the $B_s$-meson LCDA have to be taken into account as well~\cite{Beneke:2022msp}.

\section{Leading-logarithmic QCD corrections}

\label{sec:LL}

In the previous section we have identified the relevant momentum configuration for the virtual muon in the QED box diagram that is responsible for the double-logarithmic enhancement. On top of these endpoint logarithms, 
we can now include the leading-logarithmic QCD corrections to the $\bar B_s \to \gamma^*\gamma^*$ sub-process within the very same momentum configuration. A similar line of reasoning can be found, for instance, in 
Ref.~\cite{Kotsky:1997rq} for bottom-induced $h\to\gamma\gamma$ decays. Here the LO triangle diagram also contains a double-logarithmic enhancement from the endpoint configuration of the bottom-quark propagator between the two external photons. In that case, the leading double-logarithmic corrections can be included by 
dressing the off-shell $h \to b^* \bar b^*$ vertex with a standard QCD Sudakov form factor. Notice that the systematic resummation of large logarithms $\ln m_b/m_h$ for this process on the basis of RG equations in SCET has been developed only recently in Refs.~\cite{Liu:2019oav,Liu:2020wbn}.

\begin{figure}[t]

\begin{center} 
\includegraphics[width=0.49\textwidth]{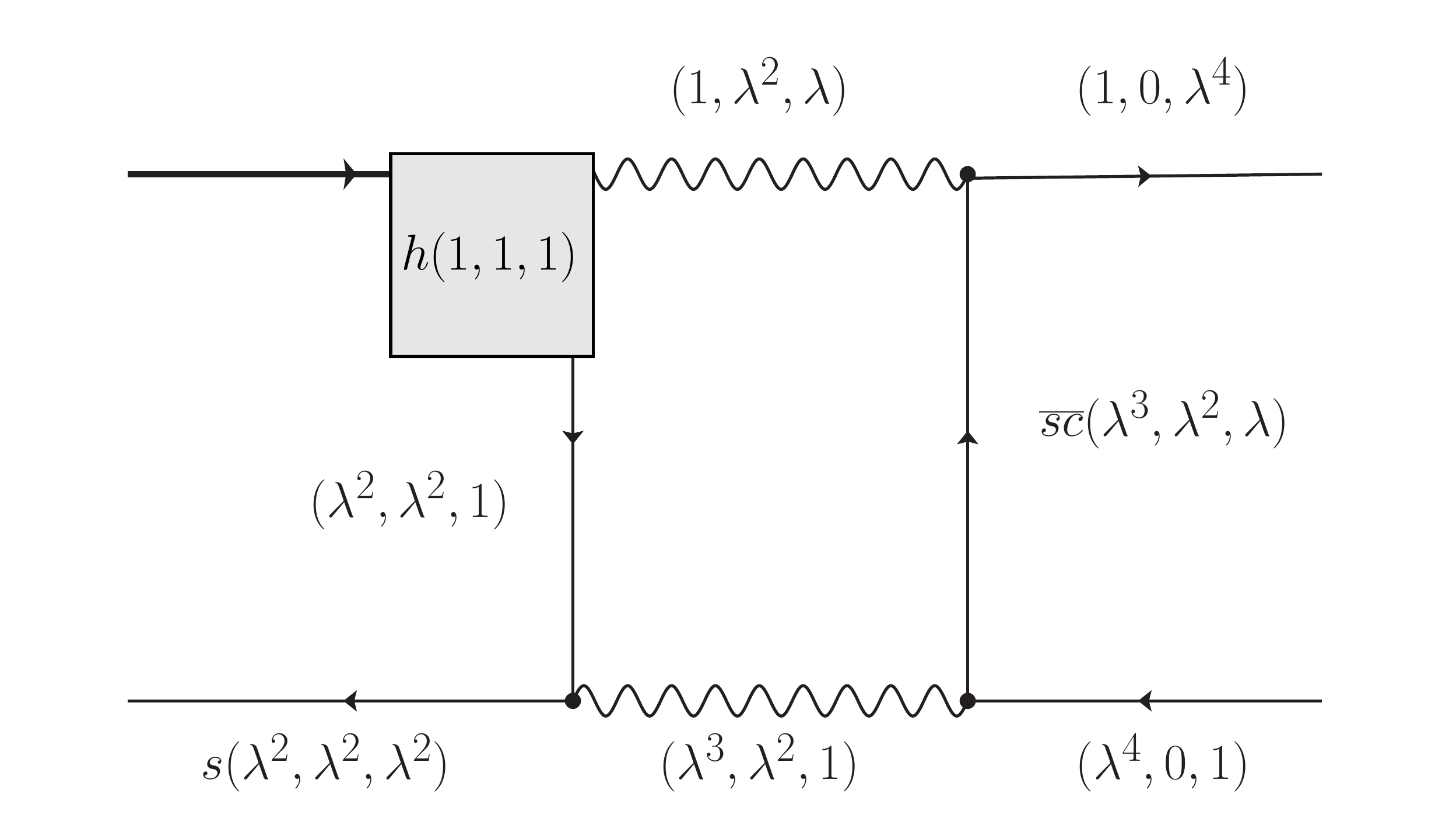}
\hfill 
\includegraphics[width=0.49\textwidth]{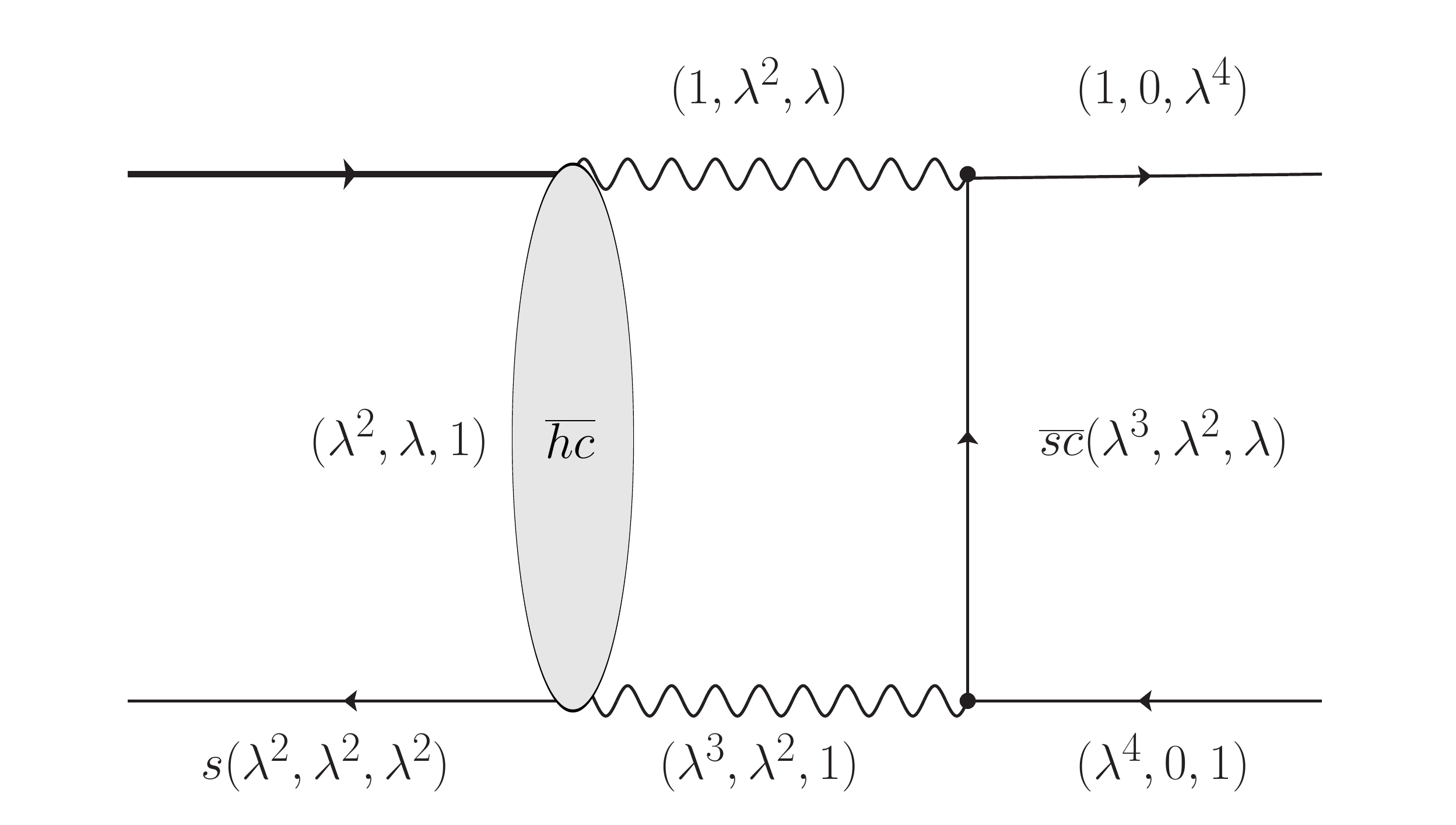}
\end{center}
\caption{\label{fig:QCDcorr} 
Illustration of short-distance QCD corrections to the anti-soft-collinear region. Left: hard correction to $b \to s\gamma^*$ vertex. Right: anti-hard-collinear corrections to spectator scattering.
}
\end{figure}

At one-loop accuracy the RG evolution is multiplicative in dual space, and -- restricting ourselves to the 
anti-soft-collinear region which generates the 
leading double-logarithmic term in Eq.~(\ref{Fdoublelog}) --  we can include the leading logarithmic radiative QCD corrections from hard, anti-hard-collinear and soft QCD effects simply via the factorization formula
\begin{eqnarray}
	{\cal F}(E,m)\Big|_{\rm LL} &=& \frac12 \, H_1(0;\mu) \,
	\int \frac{d\omega'}{\omega'} \, \rho_+(\omega';\mu) \, \ln^2 \frac{m^2 e^{2\gamma_E}}{2E \omega'} 
	\times  
{\cal \bar J}_2(1,\omega';\mu) 
 \,,
	\label{eq:LLcorr}
\end{eqnarray}
where $H_1(0,\mu)$ includes the leading-logarithmic (LL) running of the hard matching coefficient for the $b \to s \gamma^*$ vertex, and ${\cal \bar J}_2(1,\omega',\mu)$ the corresponding expression for the anti-hard-collinear jet function 
in dual space. The dual LCDA $\rho_+(\omega',\mu)$ includes the scale dependence from soft QCD modes.
This is illustrated in Fig.~\ref{fig:QCDcorr}.
The $\mu$ dependence between the individual terms in Eq.~(\ref{eq:LLcorr}) has to drop out.
We now discuss the individual terms in turn.

\subsection{Hard QCD corrections at the $b \to s\gamma^*$ vertex}

The hard corrections to the Wilson coefficients of heavy-to-light currents have been calculated in the framework of SCET, see for instance
Ref.~\cite{Bauer:2000yr}.
In the double-logarithmic approximation, we would only have to take into account the leading universal term in the corresponding solution to the RG equation,
\begin{eqnarray}
	H_1(0;\mu) &=& e^{V(\mu,\mu_h)} \left( \frac{\mu_h}{2E} \right)^{-g(\mu,\mu_h)}  \, H_1(0;\mu_h) \nonumber\\[0.2em] &\simeq &
	\exp\left[ 
	  - \frac{4\pi C_F}{b_0^2 \, \alpha_s(\mu_h)}
	  \left( \frac{1}{z_h} -1+\ln z_h \right) +\frac{2C_F}{b_0} \, \ln \frac{\mu_h}{2E} \, \ln z_h \right] H_1(0;\mu_h)
	  \,,
\end{eqnarray}
with $C_F=4/3$, 
\begin{eqnarray}
	z_h = z(\mu,\mu_h) = \frac{\alpha_s(\mu)}{\alpha_s(\mu_h)} \simeq \frac{2\pi}{2\pi + b_0 \, \alpha_s(\mu_h) \, \ln \frac{\mu}{\mu_h}}
\end{eqnarray}
and $b_0 = 11-2n_f/3$ being the leading coefficient in the QCD beta-function. 
The above formula resums all terms of order $\alpha_s^n \ln^{2n} \frac{\mu}{\mu_h}$.
The matching scale $\mu_h$ should be identified with twice
the energy of the virtual photon, which in the anti-soft-collinear region equals the muon energy:
$$
  \mu_h = (\bar n \cdot p) \simeq 2E 
$$
such that $H_1(0;\mu_h) = 1 + {\cal O}(\alpha_s)$ without logarithmically enhanced terms, and 
$$
 H_1(0;\mu) = e^{V(\mu,2E)} \, H_1(0;\mu_h=2E) \,.
$$

\subsection{Soft corrections to the $B$-meson LCDA}

The $B$-meson LCDA in dual space renormalizes multiplicatively at one-loop accuracy~\cite{Bell:2013tfa},
\begin{eqnarray}
	\rho_+(\omega',\mu) &=& e^{V(\mu,\mu_0)} \left(\frac{\mu_0 \, e^{2\gamma_E}}{\omega'} \right)^{-g(\mu,\mu_0)} \rho_+(\omega',\mu_0) 
	\,,
\end{eqnarray}
where $V(\mu,\mu_0)$ is defined analogously as before, with $\mu_h \to \mu_0$,
and 
\begin{eqnarray}
	g(\mu,\mu_0) &\simeq& - \frac{2C_F}{b_0} \, \ln z_s 
\end{eqnarray}
to leading-logarithmic accuracy,
with 
\begin{eqnarray}
	z_s = z(\mu,\mu_0)&=& \frac{\alpha_s(\mu)}{\alpha_s(\mu_0)} 
	\,.
\end{eqnarray}
The scale $\mu_0$ refers to a soft reference scale where a model/parametrization of the LCDA is defined.
Throughout this article, we set $\mu_0= 1$~GeV.

\subsection{Hard-collinear corrections to the strange-quark jet function}

Following the discussion around (2.33) in Ref.~\cite{Bell:2013tfa}, the 
leading logarithmic corrections to the jet-function describing the corrections to the 
strange-quark exchange in the anti-hard-collinear scattering process 
in dual space is again described by a multiplicative renormalization factor,
\begin{eqnarray}
  	{\cal \bar J}_2(1,\omega';\mu) &=& e^{-2V(\mu,\mu_{hc})} \left(\frac{\mu_{hc}^2 \, e^{2\gamma_E}}{2 E\omega'} \right)^{g(\mu,\mu_{hc})} {\cal \bar J}_2(1,\omega';\mu_{hc})\,,
\end{eqnarray} 
with the same  RG functions $V$ and $g$ as above.

\subsection{Net result for RG-improved endpoint contribution}

Combining all RG-factors, the dependence on the factorization scale $\mu$ drops out in the product $$
e^{V(\mu,2E)} \, {\cal J}(2E\omega',\mu) \, e^{V(\mu,\mu_0)} 
	\left( \frac{\hat\mu_0}{\omega'} \right)^{-g(\mu,\mu_0)}
	= e^{V(\mu_{hc},\mu_h)} \, e^{V(\mu_{hc},\mu_0) } \left( \frac{\hat \mu_0}{\omega'} \right)^{-g(\mu_{hc},\mu_0)} \,,
$$
as required~\footnote{We remark in passing that a very simple approximation 
for the net RG factor can be obtained by 
expanding in the coefficient $b_0$ of the QCD beta-function.
Considering, in addition, the natural relation between the individual matching scales,
$
  \mu_0 \mu_h \equiv \mu_{hc}^2 \,,
$
we obtain 
\begin{eqnarray*}
&&	H_1(0;\mu) \, {\cal \bar J}_2(1,\omega',\mu) \, e^{V(\mu,\mu_0)} 
	\left( \frac{\mu_0 \, e^{2\gamma_E}}{\omega'} \right)^{-g(\mu,\mu_0)} \Bigg|_{\mu_{hc}^2=\mu_0\mu_h}
    \\[0.3em]
	&\simeq & \exp\left[ 
	-\frac{\alpha_s(\mu_{hc}) \, C_F}{4\pi}
	\, \ln^2 \frac{\mu_0}{\mu_h}
    + \frac{\alpha_s(\mu^*) \, C_F}{\pi} \, 
    \ln \frac{\mu_0}{\mu_{hc}} \, \ln \frac{\mu_0 \, e^{2\gamma_E}}{\omega'}
	+ {\cal O}(b_0^2) \right]  
	\,, 
\end{eqnarray*}
where the scale 
$
	\mu_* = \sqrt{\mu_0 \mu_{hc}} \,
$ is determined by requiring that the ${\cal O}(b_0)$ corrections to the above approximation are zero.
The renormalization-scale for the value of $\alpha_s$ that is relevant for the $\omega'$-dependent RG effects is therefore 
given by the geometric mean of the hard-collinear and soft scale.}. Here we abbreviate $\hat \mu_0 \!=\! \mu_0 e^{2\gamma_E}$, and 
identify $\mu_h\!=\!2E$, as well as ${\cal \bar J}_2(1,\omega';\mu_{hc})\!\simeq\nobreak\! 1$.
Inserting the net leading-logarithmic RG factor into the form factor for the ${\cal O}_7$ contribution, we 
end up with 
\begin{align}
	&& {\cal F}(E,m)\Big|_{\rm LL} =
	\frac12 \, e^{V(\mu_{hc},\mu_h)} \, e^{V(\mu_{hc},\mu_0) } 
	\int \frac{d\omega'}{\omega'}\, \ln^2 \frac{m^2}{2E \hat \omega'}  \, \left( \frac{\hat \mu_0}{\omega'} \right)^{-g(\mu_{hc},\mu_0)} \rho_+(\omega',\mu_0) 
	\,, \label{res1} 
\end{align}
with $\hat \omega' = \omega' \, e^{-2\gamma_E}$.
If we define the generating function for logarithmic moments of $\rho_+(\omega',\mu_0)$ as \cite{Bell:2013tfa}
\begin{eqnarray}
    F_{[\rho_+]}(t;\mu_0,\mu_m) &=& 
    \int_0^\infty \frac{d\omega'}{\omega'} 
    \left( \frac{\hat \mu_m}{\omega'} \right)^{-t}  \rho_+(\omega',\mu_0) \,,
\end{eqnarray} 
this can also be written as ($\hat\mu_m = \mu_m e^{2\gamma_E}$)
\begin{eqnarray}
    \!\!\!\!\!\!\!\!\!\!\!\!&& {\cal F}(E,m)\Big|_{\rm LL} \!\!=
    \frac12 e^{V(\mu_{hc},\mu_h)} e^{V(\mu_{hc},\mu_0) } 
    \!\left( \frac{2E\mu_0}{m^2} \right)^{\!-g(\mu_{hc},\mu_0)} \!\!\frac{d^2}{dt^2} F_{[\rho_+]}(t+g(\mu_{hc},\mu_0);\mu_0, \frac{m^2}{2E}) \Bigg|_{t=0} .  \nonumber \\ \label{res2}
\end{eqnarray}
 The compact formulas for the form factor contain the leading double-logarithmic enhancement from the endpoint configuration of the muon propagator, supplemented by the leading-logarithmic QCD corrections in RG-improved perturbation theory. As such they still depend on the shape of the $B_s$-meson LCDA or its generating function for logarithmic moments. Thus, for 
numerical studies one has to consider models 
or generic parametrizations for $\rho_+(\omega')$ or $F_{[\rho_+]}(t)$ at the soft reference scale $\mu_0$
which is the subject of the next subsection.

\subsection{Explicit parametrization of the $B_s$-meson LCDA and numerical estimates}

\begin{figure}[t]
\begin{center}
    \includegraphics[width=0.5\textwidth]{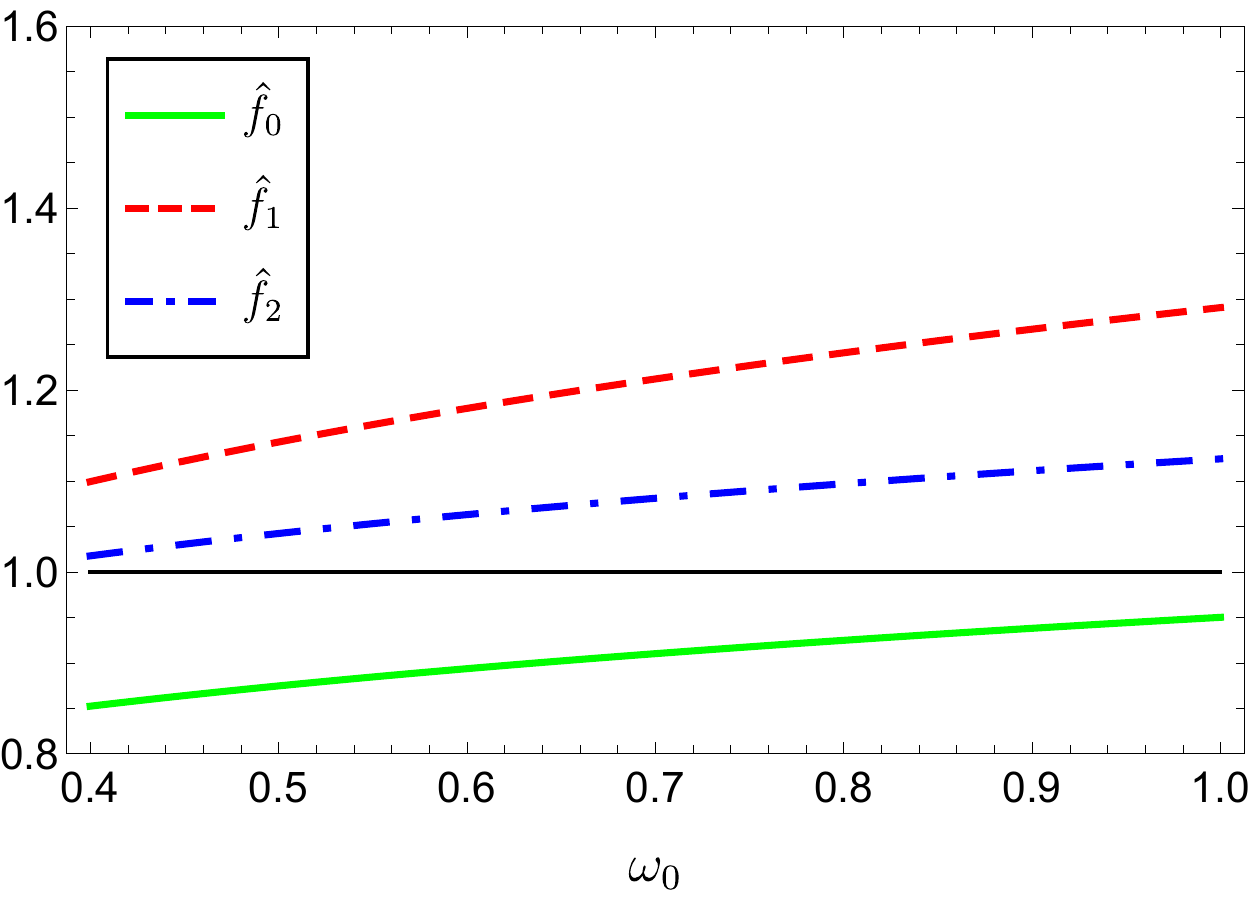}
\end{center}
\caption{\label{fig:num} 
For illustration of the RG effect, we plot the pre-factors
in front of the expansion coefficients $a_k$ defined in Eq.~(\ref{FRGexp}) as a function of the auxiliary scale $\omega_0$ and normalized to the case $g=V=0$,
which we call $\hat f_k(\omega_0)\equiv \frac{f_k(\omega_0)}{f_k(\omega_0)|_{g=V=0}}$, for $k=0,1,2$.}

\end{figure}

With the compact expression for the form factor at hand, we can use an explicit but general parametrization of the LCDA as suggested in Ref.~\cite{Feldmann:2022uok}:
\begin{eqnarray}
    \rho_+(\omega',\mu_0) &=& 
    \frac{e^{-\omega_0/\omega'}}{\omega'} \, 
    \sum_{k=0}^K \frac{(-1)^k \, a_k(\mu_0)}{1+k} \, 
    L_k^{(1)}(2\omega_0/\omega')
    \,,
    \label{eq:generic}
\end{eqnarray}
where $L_k^{(1)}$ are associated Laguerre polynomials.
The logarithmic moments of $\rho_+(\omega')$ for this parametrization can be obtained as derivatives of the generating function:
\begin{eqnarray}
    F_{[\rho_+]}(t;\mu_0,\mu_m) &=& 
    \frac{\Gamma(1-t)}{\omega_0} 
    \left( \frac{\hat \mu_m}{\omega_0} \right)^{-t} \sum_{k=0}^K a_k(\mu_0) \, {}_2F_1(-k,1+t;2;2)
    \,,
\end{eqnarray}
where the hypergeometric functions with a negative integer $-k$ as their first argument are polynomials of order $k$,
\begin{equation}
\begin{aligned}
    {}_2F_1(0, 1 + t; 2; 2)
        &= 1 \,, \cr 
    {}_2F_1(-1, 1 + t; 2; 2)
        &= -t \,, \cr
    {}_2F_1(-2, 1 + t; 2; 2)
        &= \frac{1}{3} \left(1 + 2t^2\right) \qquad \mbox{etc.}
\end{aligned}
\end{equation}
Truncating the parametrization for $\rho_+(\omega',\mu_0)$ at $K=2$, we obtain
\begin{eqnarray}
	{\cal F}(E,m)\Big|_{\rm LL} &\simeq&
	\frac{\Gamma(1-g)}{2\omega_0} \, e^{V(\mu_{hc},\mu_h)+V(\mu_{hc},\mu_0)}
	\left( \frac{\hat \mu_0}{\omega_0}\right)^{-g} 
	\cr 
	&&  {} \times \left\{ 
	\left(a_0 - g \, a_1 + \frac{1+2g^2}{3} \, a_2 \right) \left[ \left( \ln \hat \lambda_0^2+\psi(1-g)\right)^2 + \psi'(1-g)\right]
	\right.
	\cr && \qquad \left. {} + \left( 2 a_1 -\frac83 \, g\, a_2\right) \left( 
	\ln\hat\lambda_0^2  + \psi(1-g) \right) 
	+ \frac{4 a_2}{3} 
	\right\} \nonumber \\
	&\equiv& \sum\limits_{k=0}^{2} a_k \, f_k(\omega_0)
	\,,
\label{FRGexp}
\end{eqnarray}
where $\psi$ is the digamma function, 
$a_k=a_k(\mu_0)$,
$g=g(\mu_{hc},\mu_0)$, and we defined the abbreviation
$$
 \hat \lambda_0^2 \equiv \frac{m^2  e^{2\gamma_E}}{2E\omega_0} \,.
$$
To illustrate the numerical effect of the RG-improvement, we consider 
$\mu_0=1$~GeV, $\mu_h=5.3$~GeV, $\mu_{hc}=\sqrt{\mu_h \mu_0} \simeq 2.3$~GeV,
with the following values of the strong coupling at the individual scales:
$$
  \alpha_s(\mu_0)=0.49 \,, 
  \qquad \alpha_s(\mu_h)\simeq 0.21 \,,
$$
leading to
$$
    z(\mu_{hc}, \mu_h)=1.30 \,, \qquad z(\mu_{hc}, \mu_0)= 0.65 \,,
$$
for $n_f=4$, which yields 
$$
  g \simeq 0.138 \qquad \mbox{and} \qquad V(\mu_{hc}, \mu_h) \simeq -0.037 \,, \quad 
  V(\mu_{hc},\mu_0) \simeq -0.053  \, .
$$
for the relevant RG functions. From Fig.~\ref{fig:num} we see that the RG effect can be as large as $(-15,+30,+10)$\% for the contributions of the coefficients $a_{0,1,2}$, respectively, depending on the value of the auxiliary scale $\omega_0$ in the parametrization of the $B_s$-meson LCDA. Notice that the $\omega_0$-dependence of the RG effect is dominated by the factor $(\hat \mu_0/\omega_0)^{-g}$ in Eq.~(\ref{FRGexp}).

The leading-logarithmic QCD corrections to the ${\cal O}_7$ contribution have also been outlined in Ref.~\cite{Beneke:2019slt} where the additional approximation
\begin{align}
& e^{V(\mu_{\rm hc},\mu_0)} \,  \int \frac{d \omega'}{\omega'} \,
    \ln^2\frac{m^2}{2E\hat{\omega}'}
    \left(\frac{\hat{\mu}_0}{\omega'}\right)^{-g}
     \rho_+(\omega',\mu_0)
  \cr 
  & \to 
\frac{e^{V(\mu_{\rm hc},\mu_0)} \,
    \int \frac{d \omega'}{\omega'}\left(\frac{\hat{\mu}_0}{\omega'}\right)^{-g}
    \rho_+(\omega',\mu_0)
}
{
    \int \frac{d \omega'}{\omega'} \,
    \rho_+(\omega',\mu_0)
}
\cdot 
\left[
    \int \frac{d \omega'}{\omega'}
    \, \ln^2\frac{m^2}{2E\hat{\omega}'} \,
    \rho_+(\omega',\mu_0)
\right]
\cr 
&= \frac{\lambda_{B_s}(\mu_0)}{\lambda_{B_s}(\mu_{\rm hc})} \, \int \frac{d \omega'}{\omega'}
    \, \ln^2\frac{m^2}{2E\hat{\omega}'} \,
    \rho_+(\omega',\mu_0)
    \label{mom:approx}
\end{align}
has been used, which is formally valid since 
$$
  \ln^2 \frac{m^2}{2E\hat \omega'} = \ln^2 \frac{m^2}{2 E \hat\omega_0} + \mbox{sub-leading logs,}
$$
as long as $\omega_0 \sim \Lambda_{\rm QCD}$. 
For the numerical estimate of the scale-dependence of the first inverse moment $\lambda_{B_s}^{-1}$ 
of the $B_s$-meson LCDA, the authors of Ref.~\cite{Beneke:2019slt} employed the simple exponential model that corresponds to taking $a_0=1$ and $a_{k>0}=0$ in the generic parametrization~(\ref{eq:generic}). 
Adopting these additional simplifications and using otherwise the same numerical input as quoted above, we find about $(-15 \%)$ to $(-30\%)$  reduction --~depending on the value of $\omega_0$~-- from the leading-logarithmic QCD corrections to the form factor ${\cal F}$.
Up to numerical differences  related to the treatment of $\alpha_s$ and the implementation  of the RG evolution of $\lambda_{B_s}(\mu)$ \footnote{In Ref.~\cite{Beneke:2019slt} the RG factor for $\lambda_{B_s}(\mu)$ is only considered in 
fixed-order approximation, without resumming logarithms $\ln \frac{\mu_{\rm hc}}{\mu_0}$. Also note that the auxiliary parameter $\omega_0$ is identified with $\lambda_{B_s}$ in \cite{Beneke:2019slt}.}, this is in line with the numerical estimate in Ref.~\cite{Beneke:2019slt}. We also observe that without the approximation (\ref{mom:approx}) the RG effect on the $a_0$ contribution captured by the function $\hat f_0$, as illustrated by the green solid line in Fig.~\ref{fig:num},  is slightly reduced.

\section{Summary}

We have 
studied the factorization of QCD effects for the contribution of the electromagnetic dipole operator ${\cal O}_7$ to the $\bar B_s \to \mu^+\mu^-$ decay amplitude.
This arises from the non-local hadronic matrix element in Eq.~(\ref{eq:nonlocal}).
To this end, we have first 
performed a careful re-analysis of the leading QED box diagram using the method of momentum regions with two different options for an analytic regulator that has to be introduced to handle the otherwise divergent convolution integrals. In particular, we identified the relevant momentum region that is responsible for the double-logarithmic enhancement in the small ratio of soft and hard scales in the process.  
Considering the bare QCD factorization theorem that captures the different momentum regions for a given regulator, we performed the necessary subtractions to render all convolution integrals finite, such that the integrands can be renormalized in the standard manner. The so-obtained factorization theorem allows one to include radiative QCD corrections in renormalization-group improved perturbation theory.
Focusing on the kinematic configuration that is responsible for the double-logarithmic enhancement, we included the leading-logarithmic QCD corrections in a straightforward manner by taking into account the RG evolution for the hard, soft and jet function describing the relevant decay of the $\bar B_s$-meson to two (virtual) photons in QCD factorization. The leading-logarithmic RG factor takes a particularly simple form in the so-called dual space for the light-cone distribution amplitude of the $\bar B_s$-meson.
We find a compact analytic expression on the basis of a systematic parametrization of the $B_s$-meson's two-particle LCDA.
Numerically, the effect of the leading-logarithmic QCD corrections turns out to be up to $30$\% relative to the ${\cal O}_7$ contribution at fixed-order ${\cal O}(\alpha_s^0)$, in qualitative agreement with an earlier estimate in Ref.~\cite{Beneke:2019slt}.
We remind the reader that the net contribution of ${\cal O}_7$ to the total $\bar B_s \to \mu^+\mu^-$ rate is, however, small.
From the theoretical point of view it would be interesting to confirm our findings by an explicit two-loop calculation including all combinations of momentum regions, which we leave for future investigation.

\begin{acknowledgments}
We thank Guido Bell for helpful discussions and comments. TF thanks Shoji Hashimoto and the KEK Theory Group for the kind hospitality and financial support during his stay in summer 2022, where part of this work has been finalised. This research is supported by the Deutsche Forschungsgemeinschaft (DFG, German Research Foundation) under grant 396021762 -- TRR 257. Diagrams were drawn with JaxoDraw~\cite{Binosi:2003yf}.
\end{acknowledgments}

\appendix 

\section{Detailed derivation of the factorization theorem}

\label{app}

In this appendix we give a detailed derivation of the factorization theorem Eq.~(\ref{fact}), by specifying the necessary subtractions that make every single convolution integral finite. 
Starting point is the bare factorization theorem for the analytic regulator ${\cal R}_a(k)$ in Eq.~(\ref{eq:bare}),
\begin{eqnarray*}
{\cal F}(E,m) &=& \int_0^\infty \frac{d\omega}{\omega} \, \phi_+(\omega) 
\Bigg\{ \int_0^1 \frac{du}{u} \, H_1(u) \, \bar J_1(u;\omega) \,
\cr 
&& \qquad {} + \bar J_2(1,\omega) \, \int_0^1 \frac{du}{u} \, H_1(u) \, \bar C(u;\omega) 
\cr 
&&
\qquad {} + H_1(0) \, \int_0^\infty \frac{du}{u} \, \int_0^\infty \frac{d\rho}{\rho} \, S(u,\rho;\omega) \, \bar J_2(1+\rho,\omega)  \Bigg\}_{\rm bare}
\end{eqnarray*}
The procedure to perform the subtractions of endpoint divergences in the convolution integrals follows closely the analysis in Ref.~\cite{Liu:2018czl}. However, in our case the situation is somewhat simpler, because some of the functions are only needed at leading order, as long as QED corrections of order ${\cal O}(\alpha)$ are ignored.

By means of the refactorization condition (\ref{refac}) we can re-arrange the divergent convolution integral in the second line of the factorization theorem as follows,
\begin{eqnarray}
&& \bar J_2(1,\omega) \, \int_0^1 \frac{du}{u} \, H_1(u) \, \bar C(u;\omega)
\nonumber\\[0.3em]
&=& \bar J_2(1,\omega) \, H_1(0) \, \int_0^1 \frac{du}{u} \, \int_0^\infty \frac{d\rho}{\rho} \, S(u,\rho;\omega) 
\nonumber\\[0.3em]
&& {} 
+ \bar J_2(1,\omega) \, \int_0^1 \frac{du}{u} \left[ H_1(u) \, \bar C(u;\omega) - H_1(0) \left[\left[ C(u;\omega)\right]\right]
\right]
\end{eqnarray}
where the integral in the last line does not contain an endpoint divergence anymore.
Similarly, in the last term of the bare factorization, we first decompose the $\rho$-integral as \footnote{The separation of the $\rho$-integration at $\rho=1$ is convenient, but one could also cut the integral at $\rho=\rho_0 \sim {\cal O}(1)$ to obtain a more general formula.}
\begin{eqnarray}
&& H_1(0) \, \int_0^\infty \frac{du}{u} \, \int_0^\infty \frac{d\rho}{\rho} \, \bar J_2(1+\rho,\omega) \, S(u,\rho;\omega)
\nonumber\\[0.3em]
&=& H_1(0) \, \int_0^\infty \frac{du}{u} \, \int_0^1 \frac{d\rho}{\rho} \,  \bar J_2(1+\rho,\omega) \, S(u,\rho;\omega)
\nonumber\\[0.3em]
&& {} + H_1(0) \, \int_0^\infty \frac{du}{u} \, \int_1^\infty \frac{d\rho}{\rho} \,  \bar J_2(1+\rho,\omega) \, S(u,\rho;\omega)
\end{eqnarray}
where the first term on the r.h.s.\ contains the $1/\delta$ divergence, while the second term is finite in the limit $\delta \to 0$.
The former can be further decomposed 
\begin{eqnarray}
&& H_1(0) \, \int_0^\infty \frac{du}{u} \, \int_0^1 \frac{d\rho}{\rho} \,  \bar J_2(1+\rho,\omega) \, S(u,\rho;\omega)
\nonumber\\[0.3em]
&=&  H_1(0) \, \int_0^\infty \frac{du}{u} \, \int_0^1 \frac{d\rho}{\rho} \left( \bar J_2(1+\rho,\omega) -\bar J_2(1,\omega) \right) S(u,\rho;\omega)
\nonumber\\[0.3em]
&& {} +  H_1(0) \, \bar J_2(1,\omega) \, \int_0^\infty \frac{du}{u} \, \int_0^1 \frac{d\rho}{\rho} \, S(u,\rho;\omega)
\end{eqnarray}
where the $1/\delta$ divergence resides in the last term.
We can now combine the two endpoint-divergent terms as follows
\begin{eqnarray}
  && \bar J_2(1,\omega) \, H_1(0) \left( \int_0^1 \frac{du}{u}\, \int_0^\infty \frac{d\rho}{\rho} + \int_0^\infty \frac{du}{u} \, \int_0^1 \frac{d\rho}{\rho} \right)  S(u,\rho;\omega)
  \nonumber\\[0.3em]
  &=& \bar J_2(1,\omega) \, H_1(0) \left( \int_0^1 \frac{du}{u}\, \int_0^1 \frac{d\rho}{\rho} - \int_1^\infty \frac{du}{u} \, \int_1^\infty \frac{d\rho}{\rho} \right)  S(u,\rho;\omega) \Big|_{\lambda^2\to 0}
  \label{emergent}
\end{eqnarray}
where we have used that in case of analytic regulators the scaleless double-integral
\begin{eqnarray*}
\int_0^\infty \frac{du}{u} \, \int_0^\infty \frac{d\rho}{\rho} \, S(u,\rho;\omega) &=& 0 \,,
\end{eqnarray*}
vanishes. The indicated limit $\lambda^2 \to 0$ is understood to be performed after the convolutions, keeping the logarithmically enhanced terms. Notice that in the second term in brackets on the r.h.s.\ of Eq.~(\ref{emergent}) we can drop the analytic regulator and directly set $\lambda^2 \to 0$ in the soft function  (in fact, for the leading expression  $S^{(1)}(u,\rho;\omega)$, the ${\cal O}(\lambda^2)$ terms only contribute at ${\cal O}(\epsilon)$). 
This term is thus insensitive to the long-distance dynamics of the muon and thus can be combined with the first term in the factorization theorem.
Collecting all terms then leads to Eq.~(\ref{fact}) in the main text.

It is instructive, to consider the leading-order expressions for the individual contributions in the endpoint-subtracted factorization theorem (\ref{fact}).
For the first line, we obtain
\begin{eqnarray}
&& \int_0^\infty \frac{du}{u} \Bigg[ H_1^{(0)}(u) \, \bar J_1^{(1)}(u;\omega) \, \theta(1-u)
\cr && \qquad \qquad \qquad \qquad 
- H_1^{(0)}(0) \, \bar J_2^{(0)}(1,\omega) \, \theta(u-1) \, \int_1^\infty\frac{d\rho}{\rho} \, S^{(1)}(u\rho;\omega) \Bigg]_{\lambda^2 \to 0}
\cr 
&=& \left( \frac{\mu^2 e^{\gamma_E}}{2E\omega} \right)^\epsilon \,
\int_0^\infty \, \frac{du}{u^{1+\epsilon}} \Bigg[  -  \Gamma(\epsilon) \,  (1-u)^{1-\epsilon} \, H_1^{(0)}(u) \, \theta(1-u)
\cr && \qquad \qquad \qquad \qquad 	-  H_1^{(0)}(0)  \, \frac{\theta(u-1)}{ \Gamma(1-\epsilon)}\, \int_1^\infty\frac{d\rho}{\rho^{1+\epsilon} }  \Bigg] 
\cr 
&=& \left( \frac{\mu^2}{2E\omega} \right)^\epsilon \,
\int_0^1 du \Bigg[
\left( \delta(u) -\left[\frac{1-u}{u} \right]_+ \right) \frac{1}{\epsilon} 
+ 2 \, \delta(u) 
+\left[\frac{(1-u)\, \ln [u (1-u)]}{u} \right]_+ 
\Bigg]  \, H_1^{(0)}(u) 
\nonumber \\[0.25em]
&\equiv & \int_0^1 du \, \bar J_1^{\,{\rm eff(1)}}(u) \, H_1^{(0)}(u) \,.
\label{line1}
\end{eqnarray} 
We observe that the $1/\epsilon^2$ term that originally appeared in the anti-hard-collinear region from the endpoint divergence at $u\to 0$ is now cancelled by the subtraction, and the remaining contributions can be obtained from an endpoint-finite convolution with an effective anti-hard-collinear function, which is distribution-valued and reproduces the single-logarithmic term.

Similarly, for the second line of the endpoint-subtracted factorization theorem we find at leading order

\begin{eqnarray} 
&& \bar J_2^{(0)}(1,\omega) \, \int_0^1 \frac{du}{u} \Big[ H_1^{(0)}(u) \,  \bar C^{(1)}(u) - H_1^{(0)}(0) \left[\left[ C^{(1)}(u)\right]\right] \Big]
\nonumber\\[0.3em]
&=& \Gamma(\epsilon) \, 
\left(\frac{\mu^2 e^{\gamma_E}}{m^2}\right)^{\epsilon}  \int_0^1 \frac{du}{u}  \left[H_1^{(0)}(u) \, (1-u)^{1-2\epsilon}- H_1^{(0)}(0) \right]
\nonumber\\[0.3em]
&=&	\left(\frac{\mu^2}{m^2}\right)^{\epsilon}  \int_0^1 du  
\Bigg[\left( - \delta(u) + \left[\frac{1-u}{u} \right]_+ \right) \frac{1}{\epsilon}
\cr && \qquad \qquad  -
\left(2-\frac{\pi^2}{3}\right) \, \delta(u) - \left[\frac{2 (1-u) \, \ln(1-u)}{u} \right]_+\Bigg]\,  H_1^{(0)}(u) 
\nonumber\\[0.3em]
&\equiv& \bar J_2^{(0)}(1,\omega) \, \int_0^1 du \, \bar C^{\, {\rm eff(1)}}(u) \, H_1^{(0)}(u) \,,
\end{eqnarray}
which again only contains the single-logarithmic term. The associated $1/\epsilon$-divergence cancels with the one from the first line, see Eq.~(\ref{line1}), on the integrand level, with
an effective anti-collinear function which is again distribution-valued.

For the third line in the factorization theorem, we find 
\begin{eqnarray} 
&& H_1^{(0)}(0)  \, \int_0^\infty \frac{d\rho}{\rho} \left[ \bar J_2^{(0)}(1+\rho,\omega) -  
\theta(1-\rho) \, \bar J_2^{(0)}(1,\omega) \right] \, \int_0^\infty \frac{du}{u} \, S^{(1)}(u \rho;\omega)
\nonumber\\[0.3em]
&=& \left(\frac{\mu^2 e^{\gamma_E}}{2E\omega} \right)^\epsilon  \int_0^\infty \frac{d\rho}{\rho} \left[ \frac{1}{1+\rho} -  
\theta(1-\rho) \right] \int_{\lambda^2/\rho}^\infty \frac{du}{u} \, \frac{(u\rho-\lambda^2)^{-\epsilon}}{\Gamma(1-\epsilon)}
\nonumber\\[0.3em]
&=& \Gamma(\epsilon)  \left(\frac{\mu^2 e^{\gamma_E}}{m^2} \right)^\epsilon   \int_0^\infty \frac{d\rho}{\rho} \left[ \frac{1}{1+\rho} -  
\theta(1-\rho) \right] =0 \,,
\end{eqnarray} 
and therefore the soft region does not contribute at leading order after subtraction.

Finally, as already stated in Eq.~(\ref{scbarcut}), the fourth line of the subtracted 
factorization theorem reproduces the anti-soft-collinear region with cut-offs,
\begin{eqnarray*} 
 \bar J_2^{(0)}(1,\omega) \, H_1^{(0)}(0) \, \int_0^1 \frac{du}{u}\, \int_0^1 \frac{d\rho}{\rho} \,  S^{(1)}(u \rho;\omega) \Big|_{\rm \lambda^2\to 0} 
&=& \frac12 \, \ln^2 \lambda^2 \,.
\end{eqnarray*}

%\clearpage

\bibliography{references}

\end{document}